\documentclass[aps,prd,reprint]{revtex4-2}
\usepackage{amsmath}
\usepackage{multirow}
\usepackage{amssymb}
\usepackage{graphicx}
\usepackage[T1]{fontenc}
\usepackage{lmodern}

\reversemarginpar

\begin{document}

\title{Detectability Scaling Laws for Environmental Phase Modulation in Gravitational-Wave Signals}

\author{Jericho Cain}
\email{jericho.cain@gmail.com}
\affiliation{Physics Department, Portland Community College, Portland, OR, USA}
\date{\today}

\begin{abstract}

Environmental effects such as hierarchical triple motion introduce cumulative
phase modulation in gravitational-wave signals through time-dependent
line-of-sight acceleration. Whether such effects are observable depends on both
deformation strength and signal-to-noise ratio (SNR). However, this relationship
has not been quantified in a waveform-agnostic manner. Here we investigate the detectability of smooth time-warp distortions using
template-free time--frequency representations. Rather than relying on direct
strain-domain residuals, we analyze trajectory-level statistics derived from
the continuous wavelet transform, in particular the evolution of the
power-weighted frequency centroid. We show that environmental modulation can
be detected using a single-sample statistic referenced to an isolated-binary
distribution, without requiring matched templates. Across a grid of cumulative phase distortions $\Delta\phi$
and SNR, the detection performance collapses onto a single scaling parameter,
\(
\Lambda = \Delta\phi \times \mathrm{SNR}.
\)
The ROC-AUC follows an approximately sigmoid curve in $\Lambda$ with a clear transition region.
For moderate distortions ($\Delta\phi \gtrsim 3\,\mathrm{rad}$),
environmental modulation is detectable even at low SNR. For smaller distortions
($\Delta\phi \sim 1\,\mathrm{rad}$), detectability is noise-limited
and emerges only at $\mathrm{SNR} \gtrsim 20$. Within the controlled waveform family studied 
here, smooth environmental phase modulation is not generically absorbed by intrinsic waveform variability. Instead, detectability
is governed by a simple scaling between cumulative phase distortion and signal
strength. This scaling relation provides a concrete reference point for
template-free environmental searches in gravitational-wave data.

\end{abstract}

\maketitle

\section{Introduction}

Many massive stars reside in multiple systems.  Observation has shown that 
high mass stars, like O-type, are found predominately in multiple systems \cite{Sana2012} and a good chunk of those exist
in triples or higher, more so with increasing mass \cite{Shariat2025}. The systems with three stars are called 
hierarchical triple systems (HTS) and are important for gravitational wave detection.

Matched filtering using waveform templates representing binary systems have been used successfully for detection and 
parameter estimation \cite{Christensen2022}. However, HTS change gravitational waveforms systematically. Similar phase distortions can arise from other large-scale environmental accelerations, making it useful to understand their detectability in a waveform-agnostic framework. For HTS motion, 
the inner binary about a tertiary companion induces a time-dependent line-of-sight acceleration (LOSA) \cite{Gupta2020, Cardoso2021} and structured 
acceleration profiles can produce nontrivial phase evolution beyond constant-acceleration approximations \cite{Hendriks2026, Zwick2023}.
This can create phase modulation that accumulates over time producing cumulative phase drift. If in detection we assume the source is an isolated binary, but the system
is an HTS, this phase drift may be absorbed into intrinsic parameters biasing estimations of mass, chirp rate, or spin. In other words,
the LOSA signal stays inside the isolated binary envelope hidden through incorrect parameter estimation -- the effect is degenerate.

We therefore ask whether environmental phase modulation can be detected without constructing explicit environmental waveform templates.
In this work we show that the detectability of such distortions obeys a simple scaling relation governed by the product of cumulative phase deformation and signal-to-noise ratio. In particular, receiver operating characteristic (ROC) performance collapses onto a single parameter
\(
\Lambda = \Delta\phi_{env} \times \mathrm{SNR}.
\)
Using time--frequency representations computed from continuous wavelet transforms (CWT), we demonstrate this scaling by analyzing
trajectory-level statistics derived from the power-weighted frequency centroid of the signal. Detection performance remains near chance below a characteristic value of $\Lambda$, and rises rapidly above it.
In particular, phase distortions $\Delta\phi \gtrsim 3\,\mathrm{rad}$ are detectable even at low SNR, but
$\Delta\phi \sim 1\,\mathrm{rad}$ becomes observable only when $\mathrm{SNR} \gtrsim 20$.

LOSA distortions produce measurable shifts in centroid trajectories that follow predictable
SNR-$\Delta\phi$ scaling. This relation provides a physically interpretable baseline for
template-free environmental searches in current and future gravitational-wave detectors.

\section{Waveform Model and Environmental Phase Modulation}
\label{sec:waveform_model}
\subsection{Isolated Chirp Model}

To isolate the time-reparameterization effect under controlled statistical conditions we begin with a minimal inspiral-like chirp model. The same deformation behavior is also demonstrated using a post-Newtonian inspiral waveform.
\begin{equation}
h_{\mathrm{iso}}(t)
=
A(t)\cos\!\left[\phi_{\mathrm{iso}}(t)\right],
\end{equation}
where $A(t)$ is a smooth amplitude envelope and
$\phi_{\mathrm{iso}}(t)$ is a monotonically increasing phase.
The instantaneous frequency is
\begin{equation}
f_{\mathrm{iso}}(t)
=
\frac{1}{2\pi}
\frac{d\phi_{\mathrm{iso}}}{dt}.
\end{equation}

For concreteness, we employ a Gaussian-modulated chirp with
frequency evolving from $f_{\mathrm{start}}$ to $f_{\mathrm{end}}$
over the signal duration.
The precise functional form is not critical; what matters is that
$f_{\mathrm{iso}}(t)$ is smooth and increasing, mimicking the
qualitative structure of compact-binary inspiral.

\subsection{Line-of-Sight Acceleration as a Time Warp}

Environmental motion induces a detector-frame time reparameterization.
If the source experiences line-of-sight acceleration,
the observed signal becomes
\begin{equation}
h_{\mathrm{LOSA}}(t)
=
h_{\mathrm{iso}}\!\left(t + \Delta t(t)\right),
\end{equation}
where $\Delta t(t)$ is a smooth, slowly varying time shift
determined by the relative motion of the source.

In the constant-acceleration approximation,
\begin{equation}
\Delta t(t)
=
\frac{a_{\parallel}}{c}\,\frac{t^2}{2},
\end{equation}
where $a_{\parallel}$ is the line-of-sight acceleration
and $c$ is the speed of light.
More general acceleration profiles yield correspondingly smooth
$\Delta t(t)$ functions.

This deformation is purely kinematic.
The intrinsic phase evolution remains unchanged,
but it is evaluated at a shifted time coordinate.
The strain-domain difference
\begin{equation}
\Delta h(t)
=
h_{\mathrm{LOSA}}(t) - h_{\mathrm{iso}}(t)
\end{equation}
can be visually small even when the cumulative phase distortion
is significant.

\subsection{Cumulative Phase Distortion}

The observable effect of LOSA is best characterized by the
cumulative phase difference between the deformed and isolated
signals,
\begin{equation}
\Delta\phi(t)
=
\phi_{\mathrm{iso}}\!\left(t + \Delta t(t)\right)
-
\phi_{\mathrm{iso}}(t).
\end{equation}

For small time shifts,
\begin{equation}
\Delta\phi(t)
\approx
2\pi f_{\mathrm{iso}}(t)\,\Delta t(t).
\end{equation}

We define the total environmental phase distortion
\begin{equation}
\Delta\phi
=
\max_{t}
\left|
\Delta\phi(t)
\right|,
\end{equation}
which serves as a convenient one-parameter measure of deformation
strength.

In our controlled experiments, $\Delta\phi$ is varied
systematically while keeping the intrinsic waveform fixed.
For example, $\Delta\phi = 3\,\mathrm{rad}$ corresponds to an
end-of-signal time shift of order $\sim 10\,\mathrm{ms}$
for the fiducial chirp used here.

\subsection{Effect on Instantaneous Frequency}
\label{sec:effect_on_f}
Because the deformation is a time warp,
the instantaneous frequency of the observed signal becomes
\begin{equation}
f_{\mathrm{LOSA}}(t)
=
\frac{1}{2\pi}
\frac{d}{dt}
\phi_{\mathrm{iso}}\!\left(t + \Delta t(t)\right).
\end{equation}

Expanding to first order,
\begin{equation}
f_{\mathrm{LOSA}}(t)
\approx
f_{\mathrm{iso}}(t)
+
\dot{\Delta t}(t)\,
f_{\mathrm{iso}}(t)
+
\Delta t(t)\,
\dot{f}_{\mathrm{iso}}(t).
\label{eq:flosa_expansion}
\end{equation}

Thus LOSA induces a smooth deviation in the frequency trajectory,
even when the strain-domain difference appears visually subtle.
It is this trajectory-level distortion that we quantify in the
subsequent sections.

\begin{figure}[tbp]
\centering
\includegraphics[width=\linewidth]{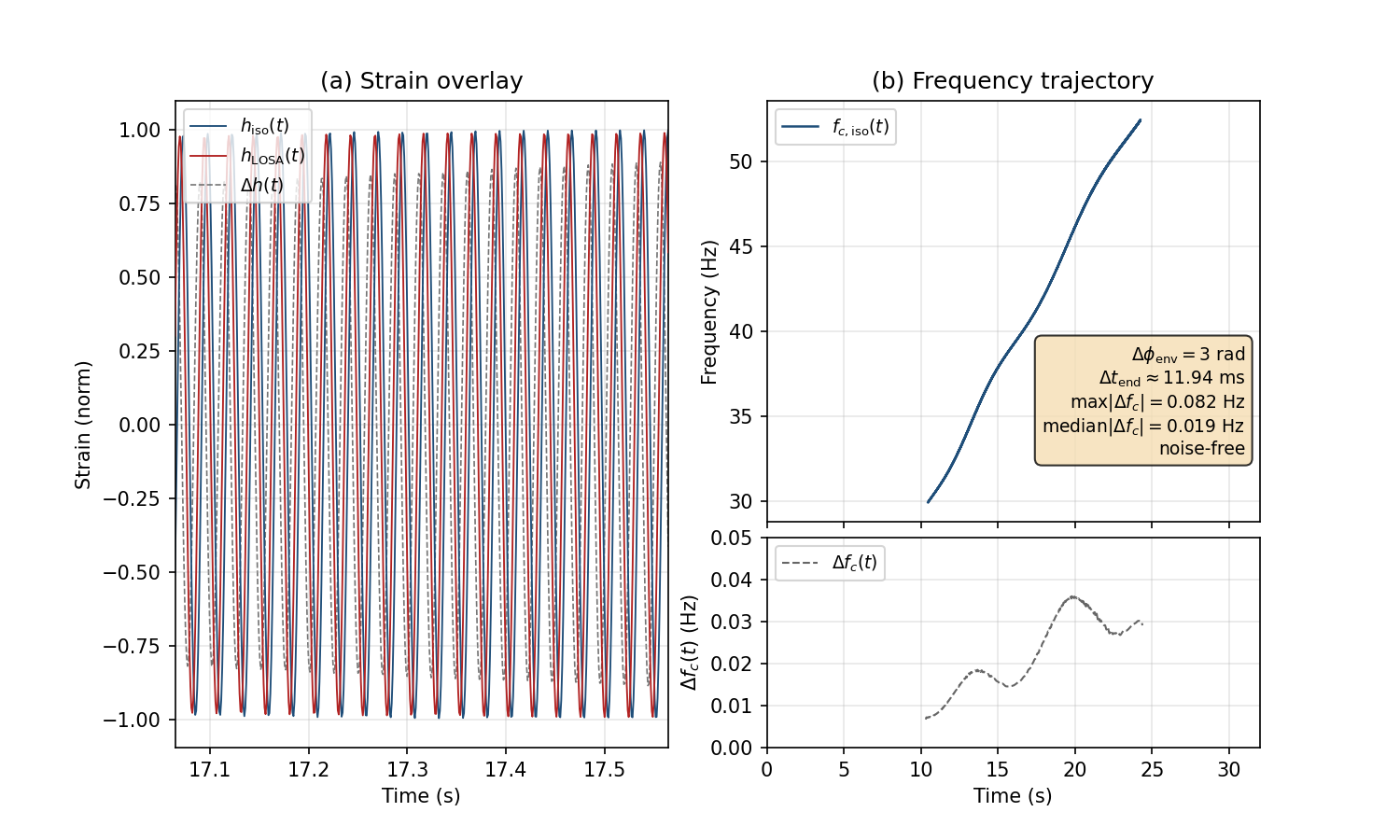}
\caption{
Environmental phase modulation for $\Delta\phi = 3\,\mathrm{rad}$ (noise-free).
(a) Zoomed strain overlay showing the isolated chirp
$h_{\mathrm{iso}}(t)$ (blue) and the LOSA-deformed signal
$h_{\mathrm{LOSA}}(t)$ (red) during the final $\sim 0.5\,\mathrm{s}$
prior to peak strain.
The dashed curve shows the strain difference
$\Delta h(t) = h_{\mathrm{LOSA}} - h_{\mathrm{iso}}$.
(b) Time--frequency centroid trajectory and its deviation,
which will be defined formally in Sec.~\ref{sec:detection_statistic}.
}
\label{fig:chirp_deformation}
\end{figure}

To illustrate the structure of the LOSA-induced deformation, we show in
Fig.~\ref{fig:chirp_deformation} a representative Gaussian-modulated chirp
example with $\Delta\phi = 3\,\mathrm{rad}$ in the noise-free limit.
Panel (a) displays a zoomed view of the final $\sim 0.5\,\mathrm{s}$
prior to peak strain in order to resolve the oscillatory phase structure.
The LOSA time reparameterization produces a cumulative multi-radian
phase slip that appears as a progressive offset between the isolated
and deformed waveforms.
Despite the several-radian cumulative distortion, the strain-domain
amplitude remains nearly unchanged.
This illustrates that LOSA acts primarily as a smooth time warp
rather than as an amplitude modulation.

\begin{figure}[tbp]
\centering
\includegraphics[width=\linewidth]{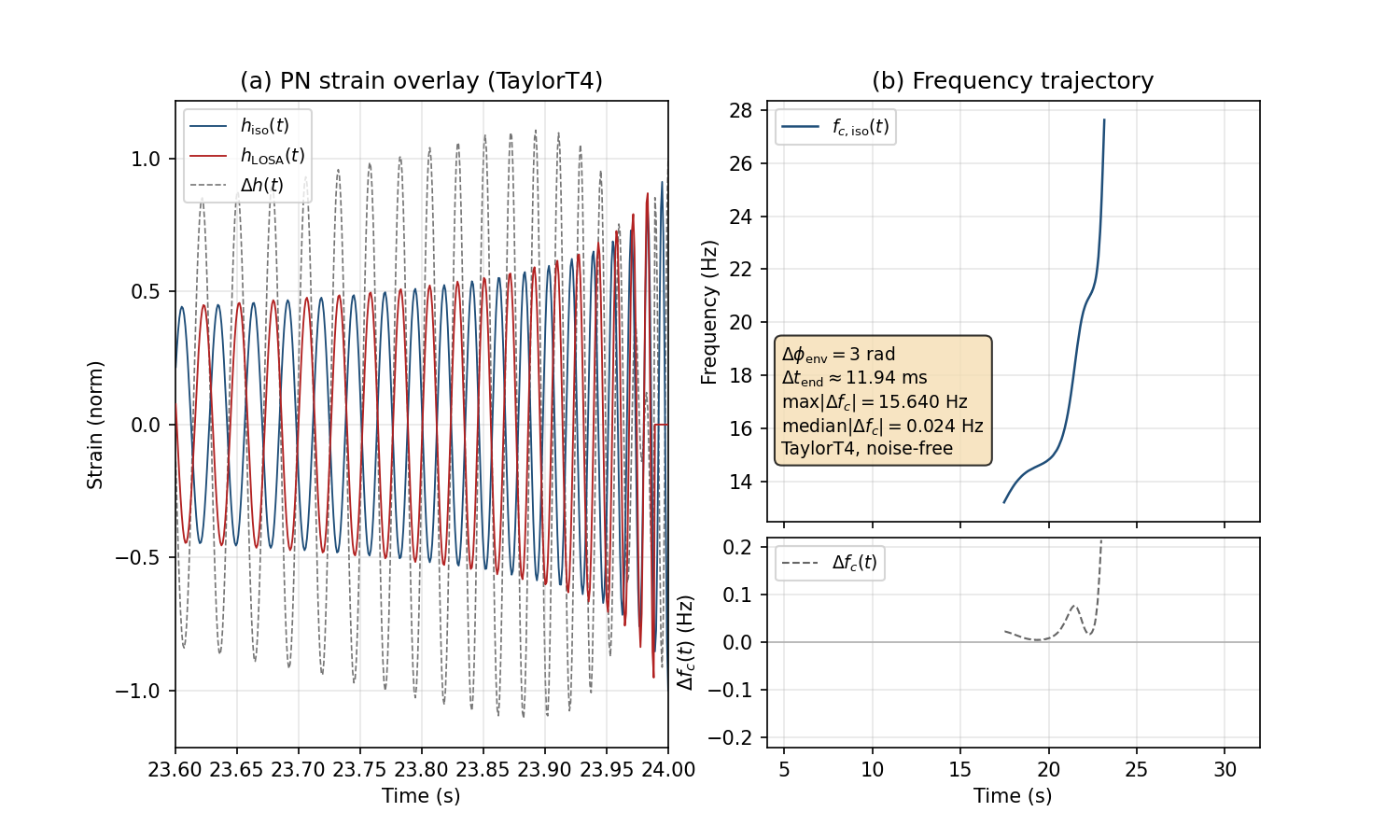}
\caption{
Environmental phase modulation applied to a post-Newtonian
(TaylorT4) inspiral waveform with
$\Delta\phi = 3\,\mathrm{rad}$ (noise-free).
(a) Zoomed strain overlay during the final $\sim 0.4\,\mathrm{s}$
before merger.
The LOSA-deformed signal (red) exhibits a cumulative phase slip
relative to the isolated waveform (blue);
the dashed curve shows the strain difference $\Delta h(t)$.
(b) Corresponding time--frequency centroid trajectory
$f_c(t)$ (top) and its deviation $\Delta f_c(t)$ (bottom).
The deformation appears as a coherent modification of the
centroid evolution, consistent with a smooth time
reparameterization rather than a constant phase offset.
}
\label{fig:pn_chirp_deformation}
\end{figure}

To demonstrate that this behavior persists for physically motivated
waveforms, we apply the LOSA time reparameterization to a
post-Newtonian inspiral generated using the TaylorT4 approximation.
Figure~\ref{fig:pn_chirp_deformation}(a) shows a representative example
with cumulative phase distortion
$\Delta\phi = 3\,\mathrm{rad}$
in the noise-free limit.
As in the Gaussian case, the deformation manifests primarily
as a progressive phase slip rather than as a large amplitude discrepancy.
This reflects the fact that LOSA does not alter the intrinsic
binary dynamics; instead, it smoothly warps the time coordinate.

For completeness, both figures also display in panel (b)
the corresponding deformation in a time--frequency representation. 
These quantities will be defined formally in Sec.~\ref{sec:detection_statistic}.

In the controlled detectability experiments that follow,
we employ the Gaussian-modulated chirp model introduced above
to isolate the deformation mechanism under well-defined
statistical conditions.
The post-Newtonian example demonstrates that the same
trajectory-level distortion arises for physically motivated
inspiral waveforms.

\section{Experimental Design}
\label{sec:design}

The goal of this study is to determine when environmental phase
modulation becomes statistically observable in noisy data.
We treat detectability as a function of two independent controls:
the cumulative environmental phase distortion $\Delta\phi$
and the signal-to-noise ratio (SNR).
Rather than relying on template matching or reconstruction error,
we evaluate a single-sample statistic defined relative to a reference
distribution of isolated signals.

\subsection{Synthetic Signal Generation}

We generate quasi-monochromatic inspiral-like chirp signals with
frequency evolution spanning $12$--$65$~Hz over a $32$~s duration.
The strain is tapered with a Gaussian envelope to suppress boundary
artifacts. Environmental modulation is introduced as a smooth
time reparameterization,
\begin{equation}
h_{\mathrm{LOSA}}(t)
=
h_{\mathrm{iso}}\!\left(t + \Delta t(t)\right),
\end{equation}
where $\Delta t(t)$ corresponds to constant line-of-sight acceleration
in the perturbative regime.

The deformation strength is characterized by the cumulative phase shift
\begin{equation}
\Delta\phi
=
\max_t
\left|
\phi_{\mathrm{LOSA}}(t)
-
\phi_{\mathrm{iso}}(t)
\right|.
\end{equation}
Unless otherwise specified, signals are embedded in stationary Gaussian
noise for statistical evaluation, while noise-free injections are used
for visual diagnostics.

Detectability is evaluated on a controlled grid of
\[
\Delta\phi \in \{0.3,\,1,\,3\}\,\mathrm{rad}
\quad\text{and}\quad
\mathrm{SNR} \in \{5,\,10,\,20,\,40\}.
\]
Additional larger distortions are used for diagnostic checks.
For each $(\Delta\phi, \mathrm{SNR})$ pair,
isolated and LOSA-modulated samples are generated independently with
distinct noise realizations. No pairing or oracle reference is used
in the primary statistical evaluation.

\subsection{Detection Statistic}
\label{sec:detection_statistic}

Time--frequency representations are computed using the continuous
wavelet transform.
From the power spectrum $P(f,t)$, we extract the power-weighted
frequency centroid
\begin{equation}
f_c(t)
=
\frac{\sum_k f_k P_{k,t}}{\sum_k P_{k,t}},
\end{equation}
restricted to gated time bins.

The time--frequency centroid $f_c(t)$ corresponds to the
upper panel (b) in
Figs.~\ref{fig:chirp_deformation}
and \ref{fig:pn_chirp_deformation}.

A reference trajectory $\mu_{\mathrm{iso}}(t)$ is constructed by
averaging $f_c(t)$ over an ensemble of isolated training signals.
For each evaluation sample (isolated or LOSA-modulated),
we compute the scalar score
\begin{equation}
S_{fc}
=
\mathrm{median}_t
\left|
f_c(t)
-
\mu_{\mathrm{iso}}(t)
\right|.
\end{equation}
This statistic measures deviation from the isolated centroid
trajectory without access to the true isolated counterpart.

Detection performance is quantified using receiver operating
characteristic (ROC) curves comparing isolated and LOSA samples.
For each grid point, the area under the ROC curve (AUROC)
is computed using $S_{fc}$ as the ranking statistic.

\subsection{Scaling Analysis}

To assess whether detectability obeys a single-parameter control,
we define the composite variable
\begin{equation}
\Lambda
=
\Delta\phi \times \mathrm{SNR}.
\label{eq:lambda}
\end{equation}
AUROC values from the full $(\Delta\phi, \mathrm{SNR})$
grid are analyzed as a function of $\Lambda$.
This allows us to test for collapse onto a universal curve
and to characterize the transition between degenerate
and separable regimes via a sigmoid fit.

\section{Results}
\label{sec:results}

\subsection{Detectability Surface in SNR and Cumulative Phase Distortion}

\begin{figure}[tbp]
    \centering
    \includegraphics[width=0.85\linewidth]{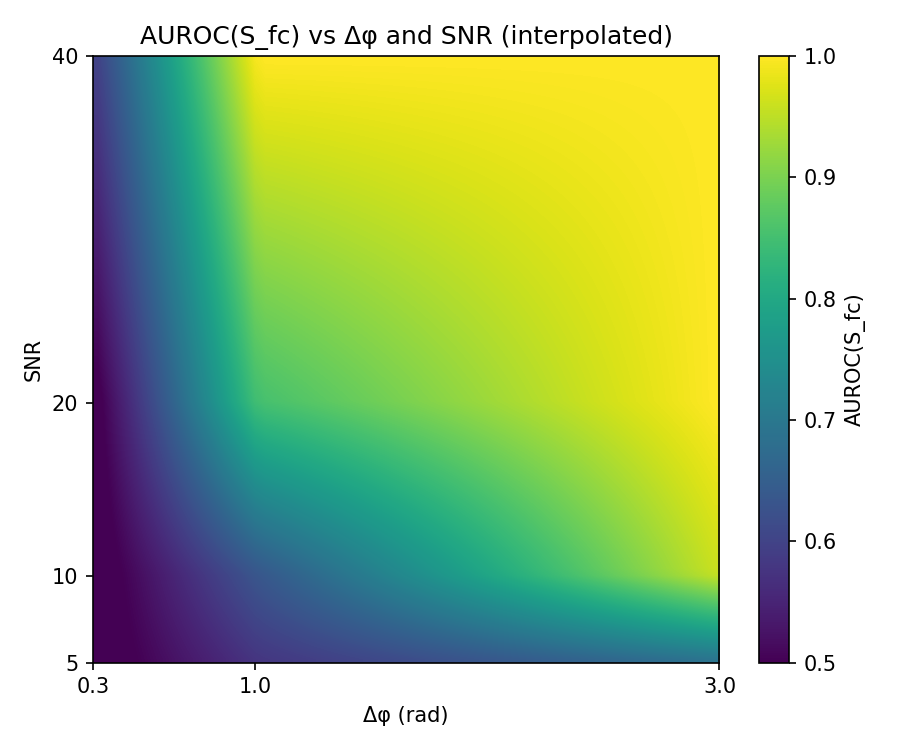}
    \caption{
    AUROC as a function of cumulative phase distortion
    $\Delta\phi$ and SNR.
    Detectability increases with both deformation strength
    and signal amplitude.
    }
    \label{fig:snr_heatmap}
\end{figure}

Figure~\ref{fig:snr_heatmap} shows AUROC as a function of
cumulative phase distortion $\Delta\phi$
and signal-to-noise ratio (SNR).
Each grid point represents an independent evaluation
using the centroid-deviation statistic $S_{fc}$.

Several features are immediately apparent:

\begin{itemize}
    \item For $\Delta\phi \approx 0.3$ rad,
    detectability remains near chance across most SNR values.
    \item For $\Delta\phi \approx 1$ rad,
    detection emerges at $\mathrm{SNR} \gtrsim 20$.
    \item For $\Delta\phi \approx 3$ rad,
    environmental modulation is detectable even at
    $\mathrm{SNR} \approx 5$.
\end{itemize}

These results indicate that detectability is jointly
controlled by deformation strength and signal amplitude.
Neither parameter alone determines observability.

\subsection{Single-Parameter Scaling and Sigmoid Transition}

\begin{figure}[tbp]
    \centering
    \includegraphics[width=0.85\linewidth]{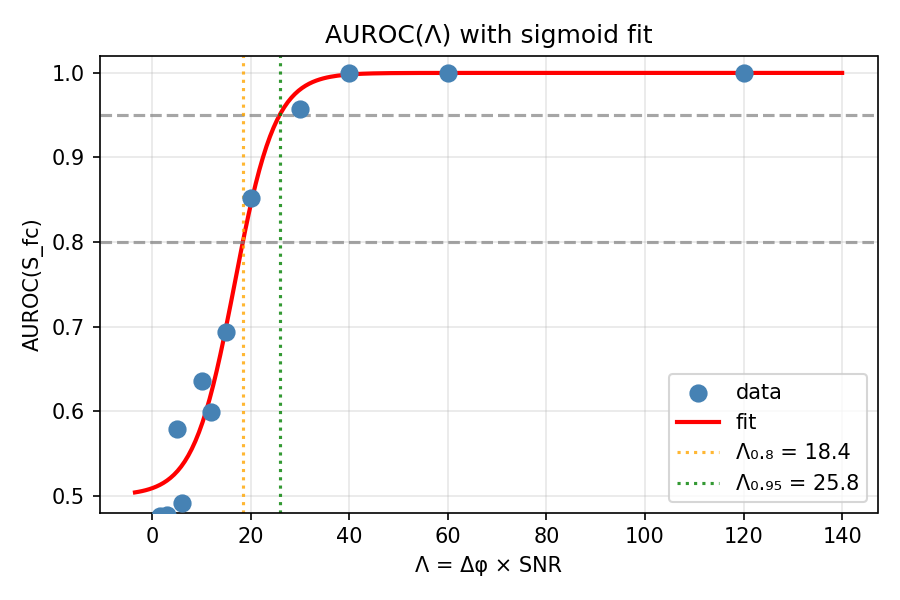}
    \caption{
    AUROC as a function of the composite parameter
    $\Lambda = \Delta\phi \times \mathrm{SNR}$.
    Points from all $(\mathrm{SNR}, \Delta\phi)$ combinations
    collapse onto a single curve, demonstrating one-parameter
    scaling of environmental detectability.
    The red curve shows a sigmoid fit to the collapsed data.
    Vertical lines indicate characteristic threshold values
    corresponding to AUROC = 0.8 and 0.95.
    }
    \label{fig:lambda_fit}
\end{figure}

Figure~\ref{fig:lambda_fit} shows AUROC plotted
against $\Lambda$ for all measured grid points. The collapse is striking: data from different
SNR and deformation combinations align along
a common monotonic trajectory.
This indicates that detectability is governed
by the product $\Delta\phi \times \mathrm{SNR}$,
rather than by either parameter independently.

To quantify the transition,
we fit the scaling curve with a sigmoid model
\begin{equation}
\mathrm{AUROC}(\Lambda)
=
0.5+\frac{0.5}{1 + \exp[-k(\Lambda - \Lambda_0)]}.
\label{eq:sigmoid}
\end{equation}

The offset of 0.5 reflects the fact that chance performance for a balanced binary classification test corresponds to $\mathrm{AUROC}=0.5$. Figure~\ref{fig:lambda_fit} shows the fitted curve
along with threshold levels corresponding to
AUROC = 0.8 and AUROC = 0.95.

The fitted parameters yield a well-defined transition scale
$\Lambda_0$.
Detection probability increases rapidly once
$\Lambda$ exceeds this threshold,
indicating a sharp transition between
noise-dominated and signal-dominated regimes.

\subsection{Time-Local Behavior of the Centroid Shift}

\begin{figure}[tbp]
    \centering
    \includegraphics[width=0.85\linewidth]{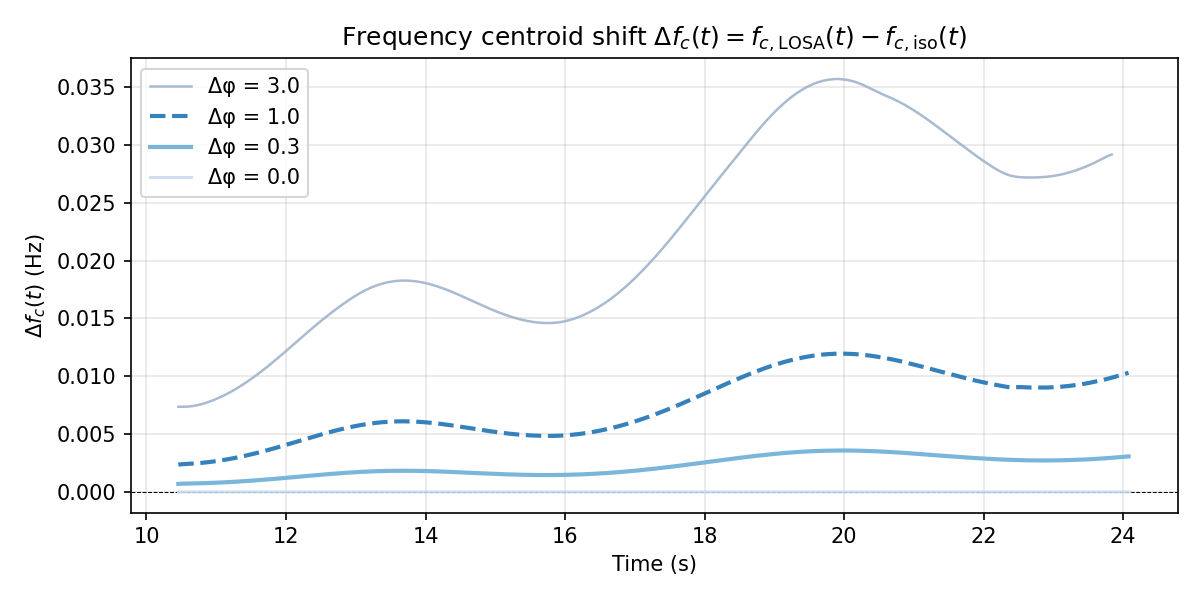}
    \caption{
    Time evolution of the centroid shift
    $\Delta f_c(t)$ for representative
    cumulative phase distortions.
    Larger $\Delta\phi$
    produces systematically larger centroid deviations,
    while preserving the characteristic smooth temporal structure.
    }
    \label{fig:delta_fc_time}
\end{figure}

To verify that the centroid-based statistic reflects
a physically meaningful deformation rather than
a numerical artifact,
we examine the explicit time evolution of the centroid difference
\begin{equation}
\Delta f_c(t)
=
f_{c,\mathrm{LOSA}}(t)
-
f_{c,\mathrm{iso}}(t),
\label{eq:delta_f}
\end{equation}
for representative deformation strengths. This quantity corresponds to the lower panel (b)
in Figs.~\ref{fig:chirp_deformation}
and \ref{fig:pn_chirp_deformation},
where a single example with
$\Delta\phi = 3\,\mathrm{rad}$
was shown.
In Fig.~\ref{fig:delta_fc_time}, we extend that visualization across multiple
cumulative phase distortions to examine the scaling behavior
of the centroid deviation.

As the cumulative phase distortion increases,
the magnitude of $\Delta f_c(t)$ grows smoothly and systematically.
The deviation exhibits coherent time structure rather than
random fluctuations, confirming that the statistic tracks
the expected time-warp deformation of the chirp trajectory.
This behavior is consistent with the interpretation developed
in Sec.~\ref{sec:waveform_model}: LOSA acts as a smooth reparameterization of time,
which manifests as a structured distortion of the
time--frequency centroid trajectory.

\subsection{Physical Interpretation}

The scaling behavior can be understood intuitively.
The cumulative phase distortion
$\Delta\phi$
sets the magnitude of the trajectory deformation,
while SNR sets the inverse noise floor
against which that deformation is measured.
Their product therefore determines the
effective signal-to-distortion ratio
in the centroid trajectory.

Within the controlled waveform family studied here, environmental phase 
modulation does not appear to be intrinsically degenerate with intrinsic 
variability in the centroid representation.
Instead, observability is governed by a
quantitative scaling relation in Eq.~(\ref{eq:lambda}).
Below the transition scale,
noise fluctuations dominate centroid deviations.
Above it, deformation-induced structure
becomes statistically resolvable.
\section{Discussion}

\subsection{Physical Interpretation of the Scaling Law}

The collapse observed in Sec.~\ref{sec:results} suggests $\Lambda$ functions 
as an effective signal-to-distortion ratio in the centroid trajectory. This scaling can be understood using simple phase-measurement arguments.
Environmental acceleration produces a deterministic cumulative phase
distortion of magnitude $\Delta\phi_{\rm env}$. In gravitational-wave
parameter estimation, the uncertainty with which waveform phase can be
measured scales approximately as $\sigma_\phi \sim 1/\mathrm{SNR}$.
Detectability therefore depends on the ratio between deformation amplitude
and measurement uncertainty, yielding an effective signal-to-distortion
parameter proportional to $\Delta\phi_{\rm env} \times \mathrm{SNR}$.
The numerical experiments presented here confirm that ROC performance
collapses onto this expected scaling.
Across the explored grid of cumulative phase distortions and
signal-to-noise ratios, AUROC values collapse onto a single
sigmoid curve when plotted against $\Lambda$.

This behavior has a direct physical interpretation.
Line-of-sight acceleration induces a smooth time reparameterization,
which in turn produces a systematic deviation in the
time--frequency trajectory of the signal.
In the perturbative regime, the magnitude of this trajectory shift
scales approximately linearly with $\Delta\phi$.
Measurement uncertainty in the centroid trajectory decreases with
increasing SNR.
Detectability is therefore controlled by the ratio between
deterministic deformation amplitude and stochastic measurement noise,
naturally leading to the product scaling in Eq.~(\ref{eq:lambda}).

The sigmoid transition reflects the crossover between
noise-dominated and deformation-dominated regimes.
For $\Lambda \ll \Lambda_0$, LOSA-induced distortions are
indistinguishable from statistical fluctuations.
For $\Lambda \gg \Lambda_0$, the deformation dominates trajectory
noise and separability approaches unity. 

The primary contribution of this work is therefore the identification of
this simple detectability scaling in $\Lambda = \Delta\phi_{\rm env}
\times \mathrm{SNR}$, while the centroid-based statistic used here serves
as a representative trajectory diagnostic for demonstrating the effect.

\subsection{Noise-Limited Versus Physics-Limited Regimes}

The detectability surface indicates that weak environmental
distortions are not fundamentally hidden by intrinsic waveform
structure.
Instead, observability is noise-limited at small
$\Delta\phi$.

For moderate distortions
($\Delta\phi \gtrsim 3\,\mathrm{rad}$),
detectability remains high even at modest SNR.
For smaller distortions
($\Delta\phi \sim 1\,\mathrm{rad}$),
separation improves rapidly with increasing SNR,
becoming reliable only once $\Lambda$ exceeds threshold.
Thus the limiting factor in the weak-deformation regime
is statistical precision rather than representational degeneracy.

\subsection{Practical Use in Detection Pipelines}

The scaling law derived here does not replace matched filtering,
but it suggests a complementary diagnostic layer for environmental searches.
After a standard binary detection, one may construct a time-frequency
representation of the recovered strain and evaluate the centroid-based
statistic $S_{fc}$ relative to an isolated reference distribution.
The inferred value of $\Lambda$
provides a quantitative criterion for whether environmental
time reparameterization is statistically resolvable.

Such a statistic could serve as a lightweight screening tool,
flagging events that warrant hierarchical or acceleration-aware
waveform modeling.
Conversely, in low-SNR regimes where $\Lambda$ is below threshold,
the analysis indicates that environmental signatures are unlikely
to be recoverable without additional information.
Beyond individual events, stacking centroid deviations across
populations may provide a statistical probe of environmental
effects even when single-event detectability is marginal.

The centroid-based statistic introduced here is also compatible with 
representation-learning frameworks in which compact-binary signals 
occupy a low-dimensional manifold in time-frequency space. If the latent representation preserves the 
time-frequency trajectory geometry, LOSA-induced deformation would correspond to a systematic displacement 
from the isolated manifold. Template-free autoencoder models trained on isolated binaries, such as \cite{Cain2026}, 
could therefore flag environmental modulation as a structured anomaly rather than generic noise. 
This suggests a natural extension in which the present scaling law is used to calibrate latent-space 
environmental sensitivity.

\subsection{Implications for Current and Future Detectors}

From the fitted sigmoid in Sec.~\ref{sec:results}, we obtain
characteristic thresholds
\[
\Lambda_{0.8} \approx 19,
\qquad
\Lambda_{0.95} \approx 27,
\]
corresponding to AUROC values of 0.8 and 0.95.

For ground-based detectors such as Advanced LIGO and Virgo,
typical binary black hole events have network
$\mathrm{SNR} \sim 8$--$20$.
At $\mathrm{SNR} \approx 10$, the condition
$\Lambda \gtrsim 20$ implies that cumulative phase distortions
of order $\Delta\phi \gtrsim 2\,\mathrm{rad}$
are required for robust detectability.
For louder events with $\mathrm{SNR} \approx 20$,
the threshold relaxes to $\Delta\phi \gtrsim 1\,\mathrm{rad}$.

Space-based detectors such as LISA operate in a different regime.
Long-duration inspirals in the milli-Hz band can accumulate very
high SNR over months to years of observation.
In that limit, even comparatively small cumulative phase distortions
may satisfy $\Lambda \gtrsim 20$.
Environmental time-warp effects that are marginal in current
ground-based observations could therefore become observable
in high-SNR space-based measurements.

\subsection{Limitations and Future Work}

The present analysis employs controlled chirp families and a
simplified constant-acceleration model in order to isolate the
time-reparameterization effect under well-defined statistical
conditions. Environmental acceleration acts primarily as a smooth
time warp, producing a cumulative phase deformation without strongly
altering the amplitude structure of the waveform. Because this
deformation is determined primarily by the accumulated phase shift
$\Delta\phi_{\rm env}$, the detectability scaling with
$\Lambda = \Delta\phi_{\rm env}\times\mathrm{SNR}$ is expected to hold
for any smoothly evolving inspiral waveform. The post-Newtonian
TaylorT4 example shown in Fig.~2 illustrates that the same
trajectory-level distortion arises for physically motivated
gravitational-wave signals.

Real hierarchical triple systems can produce time-dependent
accelerations beyond quadratic time shifts, and realistic
compact-binary waveforms contain additional structure absent from the
simplified model considered here.

Future work should extend the scaling analysis to
post-Newtonian inspiral waveforms and astrophysically motivated
acceleration profiles. Alternative trajectory statistics sensitive to
higher-order curvature in time--frequency space may further reduce
detection thresholds.

Nevertheless, within the controlled framework studied here,
environmental phase modulation exhibits a simple observability scaling
law governed by Eq.~(\ref{eq:lambda}). Detectability is controlled by
the interplay between cumulative phase distortion and signal strength,
rather than by a fundamental degeneracy with intrinsic waveform
structure.

\bibliographystyle{apsrev4-2}
\bibliography{refs}

\end{document}